\documentclass{ws-ijmpd}

\begin{document}

\markboth{N. F. Naidu and M. Govender}
{The Influence of Initial Conditions during Dissipative collapse}

%
\catchline{}{}{}{}{}
%

\title{The Influence of Initial Conditions during Dissipative collapse}

\author{N. F. Naidu\footnote{E-mail: nolene.naidu@physics.org}}
\address{Astrophysics and Cosmology Research Unit, School of Mathematical Sciences, \\
University of KwaZulu--Natal, Private Bag X54001, Durban 4000, South Africa}
\author{M. Govender\footnote{E-mail: megandhreng@dut.ac.za}}
\address{Department of Mathematics, Faculty of Applied Sciences, Durban University of Technology, Durban, 4000, South Africa.}

\maketitle

\begin{history}
\received{Day Month Year}
\revised{Day Month Year}
\end{history}

\begin{abstract}
Starting off with two distinct initially static stellar cores (i) Florides interior (constant density, vanishing radial pressure) and (ii) Wyman interior (constant density, nonvanishing radial pressure), we explore the dynamics of these two models once hydrostatic equilibrium is lost. We show that although the time of formation of horizon, evolution of the mass and proper radius are independent of the chosen initially static configurations\emph{}, there is a significant difference in the temperature profiles of the radiating bodies as the collapse proceeds.
\end{abstract}


\section{Introduction}

One of the most interesting areas in the study of general relativity and astrophysics is the end states of gravitational collapse of massive stars. Once a massive star has exhausted its thermonuclear source of energy, it begins its endless gravitational collapse. Interest in the outcome of this collapse started in 1939, when Oppenheimer and Snyder \cite{snyder} considered a spherically symmetric dust cloud which underwent continued collapse.

The first exact solution to the Einstein field equations was presented by Karl Schwarzschild in 1916 \cite{sch} which described the exterior gravitational field of a static sphere. Schwarzschild then presented the solution describing the gravitational field for the interior of a static spherically symmetric star, where the matter was assumed to be a perfect fluid with constant density. In 1951 Vaidya derived the first exact solution for the exterior gravitational field of a spherically symmetric radiating object \cite{Vaidya}. Thereafter, the junction conditions for the interior of a shear-free radiating star were derived by Santos in 1985 \cite{Santos}; and various models of radiating stars could then be studied. Hundreds of solutions incorporating shear, pressure anisotropy, charge, and several matter distributions have been found since then \cite{herrera,naidu,ivanov}.

General relativity inherently admits the existence of spacetime singularities, which are extreme regions in spacetime where densities and spacetime curvatures diverge and the theory must break down. The Cosmic Censorship Conjecture proposed by Penrose \cite{pen} in 1969, states that any reasonable matter configuration undergoing continued gravitational collapse will always form a black hole. Joshi \cite{joshi} in 2002 assessed the situation due to several models that proposed naked singularities as the final outcome of collapse. A singularity being naked implies that there exist families of future directed non-spacelike curves, which in the past terminate at the singularity. No such families exist when the collapse ends in a black hole. For black hole formation, the resultant spacetime singularity is hidden inside an event horizon of gravity, remaining unseen by external observers. For naked singularity formation, however, there is a causal connection between the region of singularity and distant observers, thus enabling communication from the superdense regions close to the singularity to distant observers. Joshi \cite{joshi} listed the physical conditions that could support the cosmic censorship conjecture and prevent the formation of a naked singularity as the end state of gravitational collapse. The conditions are: (i) a suitable energy condition must be obeyed, (ii) the collapse must develop from regular initial data, (iii) singularities from realistic collapse must be gravitationally strong (divergence of all important physical quantities such as pressure, density, curvature, etc.), (iv) the matter fields must be sufficiently general, (v) a realistic equation of state must be obeyed, (vi) all radiations from naked singularity must be infinitely red-shifted. Hence the final outcome of stellar gravitational collapse is still very much open to debate, primarily due to models that admit naked singularities \cite{harada} \cite{kudoh}. For the interior of the star, forms of the energy momentum tensor, ranging from a perfect fluid to an imperfect fluid with heat flux and anisotropic pressure have been investigated \cite{herrera,naidu}. It is well known that when a reasonable matter distribution undergoes gravitational collapse, in the absence of shear or with homogeneous density, the end result is a black hole. Shear has been identified as the factor that delays the formation of the apparent horizon, by making the final stages of collapse incoherent thereby leading to the generation of naked singularities \cite{joshi}. Wagh and Govinder \cite{wagh} showed that all known naked singularities in spherically symmetric self-similar spacetimes arise as a result of singular initial matter distributions. This is a result of the peculiarity of the coordinate transformation that takes these spacetimes into a separable form. Such examples of naked singularities are therefore of no apparent consequence to astrophysics. An interesting collapse scenario without the formation of horizon was presented by Banerjee et al. \cite{bcd1} where it was demonstrated that the non-occurrence of a horizon is due to the fact that the rate of mass loss is exactly counterbalanced by the fall of boundary radius. This model was extended to higher dimensions by Banerjee and Chaterjee who showed that the dimensionality of spacetime plays a key role in the temporal evolution of the collapsing sphere \cite{ban}.

The aim of this paper is to incorporate radiation into the Florides \cite{florides} solution and investigate the influence of the vanishing radial pressure in the initially static model on the thermal behaviour of the collapsing star. In order to determine the influence of the initial conditions of the static configuration we utilise the Wyman solution \cite{wyman} which describes a uniform density sphere with nonvanishing radial pressure. We compare the thermal behaviour of both collapsing models which had as their initial static cores described by the Wyman and Florides solutions. The dynamical evolution of these models is obtained by allowing one of the constants in the static solution to be time dependent. This approach was used previously by Kramer \cite{kramer} to model a collapsing Schwarzschild-like radiating sphere.

This paper is organised as follows: In section 2 we present the general framework which allows us to model an initially static sphere which undergoes dissipative gravitational collapse in the form of a radial heat flux. The Wyman solution which is the interior Schwarzschild solution in isotropic coordinates is introduced in section 3. The Florides solution which describes a spherically symmetric, static matter distribution with vanishing radial pressure is discussed in section 4. The thermodynamics of the collapsing models is presented in section 5. We conclude with a discussion of our results in section 6.

\section{Dissipative Collapse}

The interior spacetime of our radiating stellar model is described by a spherically symmetric shear-free line element in simultaneously comoving and isotropic coordinates of the form
\begin{equation}
ds^2 = -A_0(r)^2dt^2 + B_0^2(r)f^2(t)\left[dr^2 + r^2d\Omega^2\right],
\label{metric}
\end{equation}
where $d\Omega^2 = d\theta^2 + \sin^2{\theta}d\phi^2$, $f(t)$ encodes the dynamical nature of the model and the functions $(A_0, B_0)$ describe a static fluid solution of the Einstein field equations in isotropic coordinates. The form of the line element (\ref{metric}) has been extensively used to model shear-free collapse starting from an initial static configuration \cite{tewari,olive,olive2,olive3,gov,sharma1,sharma2,sharma3}. For our dynamical model we utilise the following stress-energy-momentum tensor
\begin{equation}\label{2}
T_{ab} = (\rho + p_T)u_au_b + p_Tg_{ab} + (p_R - p_T)\chi_a\chi_b + q_au_b + q_bu_a,
\end{equation}
where $\rho$, $p_R$, $p_T$ and $q = (q_aq^a)^{1/2}$ are the proper energy density,
radial pressure, tangential pressure and magnitude of the heat flux respectively. In comoving coordinates we have
\begin{equation} \label{vectors}
u^a = A^{-1}\delta^a_0, \hspace{1cm} \chi^a = B^{-1}\delta^a_1, \hspace{1cm} q^a = q\chi^a,\end{equation} where we identify $u^a$ as the time-like four-velocity vector  and $\chi^a$ is a unit spacelike vector along the radial direction.

The Einstein field equations for the static solution $(A_0, B_0)$ with energy momentum tensor describing a perfect fluid with anisotropic stresses can be written as
\begin{eqnarray} \label{a2}
\rho_0 &=& -\displaystyle\frac{1}{{B_0}^2}\left[
2\left(\displaystyle\frac{{B_0}^{\prime}}{B_0} \right)^{\prime} +
\left(\displaystyle\frac{{B_0}^{\prime}}{B_0} \right)^2
+ \displaystyle\frac{4}{r}\displaystyle\frac{{B_0}^{\prime}}{B_0}
\right], \label{a2a}  \\ \nonumber \\
(p_R)_0 &=&
\displaystyle\frac{1}{{B_0}^2}\left[\left(\displaystyle\frac{{B_0}^{\prime}}{B_0} \right)^2
+ 2 \displaystyle\frac{{A_0}^{\prime}}{A_0}\displaystyle\frac{{B_0}^{\prime}}{B_0}
+ \displaystyle\frac{2}{r}\left(\displaystyle\frac{{A_0}^{\prime}}{A_0} + \displaystyle\frac{{B_0}^{\prime}}{B_0}
\right)\right],  \label{a2b} \\ \nonumber \\
(p_T)_0 &=&
\displaystyle\frac{1}{{B_0}^2}\left[\displaystyle\frac{1}{r}\displaystyle\frac{{A_0}^{\prime}}{A_0} + \displaystyle\frac{1}{r}\displaystyle\frac{{B_0}^{\prime}}{{B_0}^3} + \displaystyle\frac{{A_0}^{\prime \prime}}{A_0} - \left(\displaystyle\frac{{B_0}^{\prime}}{B_0}\right)^2 + \displaystyle\frac{{B_0}^{\prime \prime}}{B_0}\right]. \label{a2c}
\end{eqnarray}
The above static solution matches with the exterior Schwarzschild spacetime across a spherical hypersurface $\Sigma$.
At this junction the pressure $p_0$ vanishes for some finite radius $
r = r_{\Sigma}$:
\[ \label{a3}
(p_0)_{\Sigma} = 0.
\]
Note that the pressure $p$ is nonzero in general because of the heat
flow. It is only for the initial static configuration that
$(p_{0})_{\Sigma} = 0$.
The Einstein field equations for the line element (\ref{metric}) and energy momentum tensor (\ref{2}) yield
\begin{eqnarray} \label{a4}
\rho &=& \displaystyle\frac{1}{f^2} \left( \rho_0 +
\displaystyle\frac{3}{{A_0}^2}{\dot{f}}^2 \right), \label{a4a} \\ \nonumber \\
p_R &=&  \displaystyle\frac{1}{f^2} \left( (p_R)_0 -
\displaystyle\frac{1}{{A_0}^2}(2f{\ddot{f}} + {\dot{f}}^2) \right),
\label{a4b} \\ \nonumber \\
p_T &=& \displaystyle\frac{1}{f^2}\left( (p_T)_0 -
\displaystyle\frac{1}{{A_0}^2}(2f{\ddot{f}} + {\dot{f}}^2) \right),
\label{a4c} \\ \nonumber \\
q &=&
-\displaystyle\frac{2{A_0}^{\prime}}{{A_0}^2{B_0}^2}\displaystyle\frac{\dot{f}}{f^3},
\label{a4d}
\end{eqnarray}
where $f(t)$ is determined by the junction condition.
In order to obtain a complete model of a collapsing star we need to invoke the junction conditions which allow for the smooth matching of the interior spacetime to the exterior spacetime. Since the star is radiating energy in the form of a radial heat flux, the exterior spacetime is no longer empty and is described by Vaidya's outgoing solution
\cite{Vaidya}
\begin{equation}
ds_{+}^2 = -\left(1-\frac{2m(v)}{\sf r}\right)dv^2 - 2dvd{\sf r} + {\sf r}^2 d \Omega^2 , \label{Vm}
\end{equation}
where the mass function $m(v)$ is a function of the retarded time $v$. The junction conditions for the matching of the line element (\ref{metric}) and the exterior Vaidya spacetime (\ref{Vm}) across a time-like hypersurface are
\begin{eqnarray} \label{aa5}
(rB_{0}f)_{\Sigma} &=& {\sf{r}}_{\Sigma}, \label{aa5a}  \\  \nonumber \\
(p_R)_{\Sigma} &=& (qB_{0}f)_{\Sigma},  \label{aa5b}  \\ \nonumber \\
\left( r(rB_{0}f)_r \right)_{\Sigma}
&=&
(\dot{v}({\sf{r}} - 2m) + \dot{{\sf{r}}}{\sf{r}}
)_{\Sigma},  \label{aa5c} \\  \nonumber \\
m(v) &=& \left( \displaystyle\frac{r^3{B_0}^3f{\dot{f}}^2}{2{A_0}^2}
- r^2{B_0^{\prime}}f - \displaystyle\frac{r^3}{2}
\displaystyle\frac{{{B_0}^{\prime}}^2f}{B_0} \right)_{\Sigma}, \label{aa5d} \\ \nonumber \\
\left( \displaystyle\frac{dm}{dv} \right)_{\Sigma} &=&
\left(-\displaystyle\frac{p}{2}\displaystyle\frac{{\sf{r}}^2}{{\dot{v}}^2}
\right)_{\Sigma}, \label{aa5e} \\ \nonumber \\
{\dot{v}}_{\Sigma} &=& \left( 1 +
r\displaystyle\frac{{B_0}^{\prime}}{B_0} +
r\displaystyle\frac{B_0}{A_0}\dot{f} \right)^{-1}_{\Sigma}.
\label{aa5f}
\end{eqnarray}
Using the junction condition (\ref{aa5b}) together with (\ref{a4b}) and
(\ref{a4c}), and taking into account $(p_{0})_{\Sigma} = 0$, we obtain
\begin{equation} \label{a5}
2f{\ddot{f}} + {\dot{f}}^2 - 2a{\dot{f}} = 0,
\end{equation}
which governs the behaviour of $f$. The constant
\begin{equation}  \label{a6}
a = \left(\displaystyle\frac{{A_0}^{\prime}}{B_0} \right)_{\Sigma},
\end{equation}
is positive because the static solution $(A_{0}, B_{0})$
must match with the
\emph{}exterior Schwarzschild metric.
A first integral of (\ref{a5}) is given by
\begin{equation} \label{a7}
\dot{f} = -2a\left(\displaystyle\frac{b}{\sqrt{f}} - 1\right),
\end{equation}
where the constant of integration is $-2ab$.
Note that $p_{\Sigma}$ is nonnegative so on utilising the result
$(p_{0})_{\Sigma} = 0$, (\ref{a4b}) and (\ref{a5}) we obtain
\[
\dot{f}(t) \leq  0.
\]
This implies that the only possible evolution of the system is
contraction. Then on using (\ref{a7}) and the fact that $f(t)$ is positive we
have
\[
0 \leq f(t) \leq b^2.
\]
Integrating (\ref{a7}) we obtain
\begin{equation} \label{a8}
t = \displaystyle\frac{1}{a}
\left[\displaystyle\frac{1}{2}f + b\sqrt{f} + b^2\ln{\left(1 -
\displaystyle\frac{\sqrt{f}}{b} \right)} \right],
\end{equation}
where the constant of integration has been absorbed by means of
rescaling the time coordinate as $
t \rightarrow t + {\mbox{constant}}$.

An analysis of (\ref{a8}) shows that the function $f(t)$ decreases
monotonically from its value $b^2$ at $ t = -\infty$ to zero at $ t = 0 $
where a physical singularity is encountered.
Physically this implies that the collapse starts at $ t = -\infty$ from
a static perfect fluid sphere with its interior described by the
solution $(A_{0}, b^2B_{0}) $ and whose energy density and pressure are
given by (\ref{a2}) provided that the right hand side of these equations
are divided by a factor of $b^2$. For convenience we set
\[
b = 1.
\]
The initial mass of the static sphere is given by
(\ref{aa5d}):
\[ \label{a9}
m_0 = -\left(r^2{B_0}^{\prime} +
r^3\displaystyle\frac{{{B_0}^{\prime}}^2}{2B_0}\right)_{\Sigma},
\]
where primes denote differentiation with respect to $r$. Its initial `luminosity radius' is given by
\[ \label{a10}
{\sf {r}}_0 = (rB_0)_{\Sigma}.
\]
At $t = -\infty$ the exterior spacetime is described by the vacuum
Schwarzschild solution in isotropic coordinates, that is
\[
ds^2 = -\left(\displaystyle\frac{1 - \displaystyle\frac{m_{0}}{2r}}{1 +
\displaystyle\frac{m_{0}}{2r}}\right)^2dt^2 + \left(1 + \frac{m_{0}}{2r}\right)^4
\left[dr^2 +
r^2 d\Omega^2\right].
\]
We match the static perfect fluid solution to the exterior Schwarzschild
solution in isotropic coordinates.
Continuity of the metric functions for the initially static sphere yields
\begin{eqnarray} \label{a11}
{A_0}^{\prime}_{\Sigma} &=& \displaystyle\frac{m_0}{{{\sf
{r}}_0}^2}\left(1 + \displaystyle\frac{m_0}{2r_{\Sigma}}\right)^2,
\label{a11a} \\ \nonumber \\
{B_0}_{\Sigma} &=& \left(1 + \displaystyle\frac{m_0}{2r_{\Sigma}}\right)^2.
\label{a11b}
\end{eqnarray}
We can rewrite $a = \left(\displaystyle\frac{{A_{0}}^{\prime}}{B_{0}}\right)_{\Sigma}$ in terms of the initial quantities $m_0$ and
${\sf {r}}_0$:
\[
a = \displaystyle\frac{m_0}{{{\sf{r}}_0}^2}.
\]
Thus $a$ is determined by the mass $m_{0}$ of the initial static
configuration and the initial luminosity radius ${\sf {r}}_{0}$. A detailed discussion of the matching conditions for both the initially static model and the dynamical model was provided by Bonnor et al.\cite{bone}

The Einstein field equations in terms of $f$ can be written as
\begin{eqnarray}
\rho &=& \displaystyle\frac{1}{f^2} \left[{\rho}_0 + \displaystyle\frac{12 a^2}{{{A}_0}^2}\left(\displaystyle\frac{1}{\sqrt{f}}-1\right)^2\right], \label{efef1}\\ \nonumber \\
{p}_r &=& \displaystyle\frac{1}{f^2} \left[(p_R)_0 - \displaystyle\frac{4 a^2}{{{A}_0}^2}\left(\displaystyle\frac{1}{\sqrt{f}}-1\right)\right], \label{efef2}\\ \nonumber \\
{p}_T &=& \displaystyle\frac{1}{f^2} \left[(p_T)_0 - \displaystyle\frac{4 a^2}{{{A}_0}^2}\left(\displaystyle\frac{1}{\sqrt{f}}-1\right)\right], \label{efef3}\\ \nonumber \\
q &=& -\displaystyle\frac{4 a {A_0}^{\prime}}{{A_0}^2 {B_0}^2} \left(\displaystyle\frac{1}{f^3}\right)\left(\displaystyle\frac{1}{\sqrt{f}}-1\right). \label{efef4}
\end{eqnarray}
In the sections that follow we will choose two initially static configurations: (i) Florides solution which describes a uniform density sphere in which the radial pressure vanishes at each interior point of the matter distribution, (ii) Wyman solution which is the constant density interior Schwarzschild solution in comoving and isotropic coordinates to study the subsequent gravitational collapse of these bodies in the presence of dissipation.

\section{Wyman Solution}

The Wyman solution is actually the interior Schwarzschild solution in comoving and isotropic coordinates. The interior matter distribution is composed of a core with uniform density. The exterior spacetime is described by the vacuum Schwarzschild solution in isotropic coordinates. The gravitational potentials for the  Wyman solution are given by
\begin{eqnarray} \label{u}
B_0=\frac{2\sqrt{3}~c}{\sqrt{\mu}(1+c^2r^2)}, \label{u2}
A_0=d \left[\frac{1+c^{2}r^2}{1-c^{2}r^2}\right]^{\frac{\mu
(1+3\alpha)-3\beta}{2 \mu}} \label{u3},
\end{eqnarray}
where $\alpha$, $\beta$, $c$ and $\mu$ are arbitrary constants and $d$ is an integration constant determined by the matching conditions.

The Einstein field equations for the time-dependent Wyman solution are
\begin{eqnarray}
\rho &=& \frac{1}{f^2} \left[\mu +\frac{12a^2}{d^4} \left( \frac{1 - c^2r^2}{1 + c^2r^2}\right)^{2y} \left( \frac{1}{\sqrt{f}} -1\right)^2 \right], \label{wf1}\\ \nonumber \\
p_R &=& \frac{1}{f^2} \left[-\beta + \alpha \mu - \frac{4 a^2}{d^2} \left( \frac{1 - c^2r^2}{1 + c^2r^2}\right)^{2y} \left( \frac{1}{\sqrt{f}} - 1 \right) \right],\label{wf2}\\ \nonumber \\
p_T &=& \frac{1}{f^2} \left[ \frac{3 r^2 \beta^2 - \beta \mu \left(1+r^4 + r^2 (2+6\alpha) \right) + \mu^2 \left(\alpha + r^2 \left( 1 + \alpha \left( 2 + r^2 + 3 \alpha \right) \right) \right)}{\mu (r^2 -1)^2}\right] \nonumber \\
&&+\left[\frac{4a^2}{f^2d^2} \left( \frac{1 - c^2r^2}{1 + c^2r^2}\right)^{2y} \left(\frac{1}{\sqrt{f}} -1\right) \right],\label{wf3}\\
q &=& \frac{4 a r (-1 + \sqrt{f})(-3 \beta + \mu + 3 \alpha \mu)\left( \frac{1+r^2}{1-r^2} \right)^{\frac{3 \beta + \mu -3\alpha \mu}{2 \mu}}}{\sqrt{3} d \sqrt{\mu} f^{\frac{5}{2}} (1+r^2)}, \label{wf4}
\end{eqnarray}
where $y = \frac{\mu(1+3\alpha) -3\beta}{2\mu}$ is a constant.
The interior Schwarzschild solution was utilised by Kramer\cite{kramer} to model a collapsing, radiating star. Starting off with the static, interior Schwarzschild solution in noncomoving coordinates, Kramer transforms this solution to isotropic coordinates. In order to obtain a nonstatic solution, Kramer allowed the mass function in the transformed interior Schwarzschild solution to become time dependent. This nonstatic generalised Schwarzschild interior is matched to the exterior Vaidya spacetime. In our approach, the time dependence which encodes the dynamical nature of the collapse resides in $f(t)$. From Figure 1. we observe that the density is a monotonically decreasing function of the radial coordinate. Figure 2. shows the trend in the radial and transverse pressures. The radial pressure is highest at the centre of the star (hottest region) and drops off monotonically towards the cooler surface layers. The anisotropy factor $\Delta = p_T - p_R$ is displayed in Figure 4. and we observe that it is positive everywhere at each interior point of the star. For our model $\Delta>0$ which confirms that the anisotropic factor is repulsive in nature and according to Gokhroo and Mehra it helps to construct more compact objects \cite{gokhroo}. Moreover, at the center of the star $\Delta$ vanishes which is also satisfied by our model.

\section{Florides Solution}

An Einstein cluster consists of a large number of particles that move in randomly oriented concentric circular orbits under their own gravitation. Einstein wrote his 1939 paper to answer the question: can physical models exhibit a Schwarzschild singularity? \cite{einstein} The particles in Einstein's model move in circular orbits with the only influence being gravity. The other simplifications are that there is no radiation emitted by the particles and the orbits of the particles are non-overlapping (there are no particle interactions, per se). Though the initial model was crude, it served a purpose: it was an easy to imagine a cluster with some physical properties and did not exhibit any singularities. Einstein showed that equilibrium is still achievable with vanishing pressure, by making use of rotation to balance the gravitational force.  The Florides solution \cite{florides} describes a field with vanishing radial pressure but nonzero tangential pressure, and may thus be interpreted as the field inside an Einstein cluster.  The solution is for a static, spherically symmetric distribution of dust, that has vanishing radial pressure but nonzero tangential pressure. The tangential pressure is used instead of the rotation Einstein used, to counteract gravity thereby achieving equilibrium. Florides has a positive, isotropic tangential pressure on the surface of the dust sphere. The radius can approach the Schwarzschild singularity value of $2m$ and get arbitrarily close to it.

Florides noted that a spherical distribution of dust would not be in equilibrium by itself: it would collapse under its own gravity. Instead of following the methodology Einstein used, i.e. taking $T^{0}_{0}$ as the only non-vanishing component of the energy tensor, Florides chose
\begin{equation}
T^{1}_{1} = T^{2}_{2} = \frac{m \rho(r)}{2r(1-2m /r)},
\label{ts}
\end{equation}
\begin{equation}
T^{0}_{0} = -\rho(r),
\end{equation}
as the only non-vanishing components, where $\rho$ is the proper density and
\begin{equation}
m = m (r) = 4 \pi \int_0^r \! \rho(\omega)\omega^2 \mathrm{d}\omega,
\end{equation}
where $\omega$ is an integration variable. The metric becomes
\begin{equation}
ds^2 = -\left(1-\frac{2m}{a}\right) \exp \left[\int_a^r \! \frac{2 m(r) dr}{r^2 (1-2m(r) /r)}\right] \mathrm{d} t^2 + \frac{dr^2}{(1-2 m(r) /r)} + r^2 d\Omega^2,
\end{equation}
with the radial pressure emerging as zero everywhere, the proper density staying positive, and the tangential stress being both positive and isotropic at any point of the surface which is given by (\ref{ts}).
If the density of the Einstein cluster is given by
\begin{equation}
\rho^{~} = \rho(1-3 m(r)/r)(1-2m(r)/r)^{-1},
\end{equation}
then the Florides solution describes the field inside that Einstein cluster. One further restriction must be made on the radius, i.e. $r>3m(r)$ for all $r \leq a$ and $a>3m$ where $a$ is the radius of the sphere at the boundary and $m$ is the gravitational mass contained within a sphere of radius $r$.
Here we have constant density and $(p_R)_0 = 0$ for all $r$.
Following Thirukkanesh et al. \cite{thiru} we employ the following forms of the gravitational potentials for the Florides solution
\begin{eqnarray}
A_0 &=& \frac{d^2\sqrt{1 + c^2r^2}}{\sqrt{1 - c^2r^2}}, \label{f1}\\ \nonumber \\
B_0 &=& \frac{2\sqrt{3}~c}{\sqrt{\mu}(1+c^2r^2)}, \label{f2}
\end{eqnarray}
where $c$ and $\mu$ are arbitrary constants.
The thermodynamical variables for Florides solution are given by
\begin{eqnarray}
\rho &=& \frac{1}{f^2} \left[\mu +\frac{12a^2 (1 - c^2 r^2)}{d^4 (1 + c^2r^2)} \left( \frac{1}{\sqrt{f}} -1\right)^2 \right], \label{rf1}\\ \nonumber \\
p_R &=& \frac{1}{f^2} \left[- \frac{4 a^2 (1 - c^2r^2)}{d^4 (1 + c^2r^2)} \left( \frac{1}{\sqrt{f}} - 1 \right) \right],\label{rf2}\\ \nonumber \\
p_T &=& \frac{1}{f^2} \left[ \frac{\mu r^2}{(-1 + r^2)^2} + \frac{4 a^2 (1 - c^2r^2)}{d^4 (1 + c^2r^2)}\left( \frac{1}{\sqrt{f}} - 1 \right) \right],\label{rf3}\\ \nonumber \\
q &=& \frac{4 a r \sqrt{\mu} (-1 + \sqrt{f})}{d^2 f^{\frac{5}{2}} \sqrt{3 - 3r^4}}. \label{rf4}
\end{eqnarray}
We should point out that since the radial pressure for the static Florides solution is zero everywhere inside the star, it follows that the temporal evolution equation (\ref{a5}) holds true everywhere inside the collapsing body as well. This is different for the dynamical Wyman model where (\ref{a5}) only holds at the boundary of the star.
Figure 4. shows that the density starts off at a high value at the stellar center and decreases monotonically outwards towards the surface. The radial pressure is a monotonically decreasing function of the radial coordinate $r$ as shown in Figure 5. The behaviour exhibited by the energy density and radial pressure supports the characteristics of a hot core dissipating energy to the exterior via a radial heat flux. The anisotropy factor is positive (Figure 6.), indicative of repulsion within the core. This repulsion can slow down the collapse rate.

\section{Thermodynamics}

Heat flow in relativistic astrophysics has been widely studied within the context of extended irreversible thermodynamics. Various studies have demonstrated that relaxational effects lead to higher core temperatures during dissipative collapse. In order to distinguish between the two initially static models we make use of a truncated causal heat transport equation of Maxwell-Cattaneo form\cite{kevin1}
\begin{equation}{\tau} h_a{}^b {\dot q}_{b} + q_a =
-\kappa(h_a{}^b \nabla_b T + T {\dot u}_a), \label{cmc}
\end{equation}
where we have assumed no viscous-heat coupling of the thermodynamical fluxes. The truncated causal heat transport equation addresses several of the pathologies that are intrinsic to the Eckart transport equation. However, the truncation implies disregarding nonlinear terms which may be significant at higher temperatures or during the latter stages of collapse\cite{gov}. We are able to calculate the causal temperature profiles for both the Wyman and Florides models for the case of constant collision time by adopting a power-law dependence for the thermal conductivity and relaxation time:
\begin{equation}
\kappa =\gamma T^3{\tau}_{\rm c}, \hspace{2cm} \tau_{\rm c}
=\left({\phi\over\gamma}\right) T^{-\sigma}, \label{a28}\,
\end{equation}
where $\phi \geq 0$, $\gamma \geq 0$ and $\sigma \geq 0$ are
constants.
We also assume that the relaxation time is directly proportional to the mean collision time \begin{equation}
\tau =\left({\psi \gamma \over \phi}\right) \tau_{\rm c},
\label{a30}\,
\end{equation}
where $\psi$ ($\geq 0$) is a constant.
Figures 7. and 8. clearly show that the casual temperature is greater than the Eckart temperature within the fluid distribution for the respective models. Furthermore, cooling is enhanced at the surface layers where heat generation is much lower compared to the central dense regions of the star. Figure 7. highlights an interesting observation regarding the Eckart temperature distribution of the Wyman model (solid line) and Florides model (dashed line). The Eckart temperature for the Wyman model is lower than its counterpart for the Florides model. This is true for each interior point of the respective models. The same conclusions hold true for the causal temperature profiles (Figure 8.). This behaviour in the temperature profile is expected as the nonzero radial pressure present in the Wyman model slows down the collapse thus hindering heat generation within the core. The interplay between the tangential pressure and radial pressure in the Wyman model results in a high anisotropy factor which is responsible for a repulsion effect within the stellar core. This repulsion slows down the collapse leading to lower heat generation. The anisotropy factor in the Florides model is three orders of magnitude smaller than that in the Wyman model. The repulsion is significantly smaller thus allowing for a more efficient collapse and larger heat dissipation throughout the core.

\section{Conclusion}
We have utilised two initially static matter distributions to model dissipative collapse. Both models start off with initially static cores of uniform density, same initial mass and initial radii,  the crucial difference being that the Florides model has vanishing radial pressure at each interior point of the stellar configuration while the Wyman solution has a nonzero radial pressure throughout the stellar interior. It is interesting to note that the presence of the radial pressure does not affect the time of formation of the horizon which depends only on the initial mass and radius of the initial static configuration. Also worth mentioning is the effect of the radial pressure on the temperature profiles in both the causal and noncausal theories. The radial pressure (through the anisotropy parameter) slows down the collapse rate thus reducing the amount of heat being generated within the core. This results in a lower core temperature when radial pressure is present. By employing a truncated causal heat transport equation we showed that the presence of the initial radial pressure within the static core is further enhanced by relaxational effects. We believe that this is the first time that such an effect on the temperature profile of the collapsing body has been demonstrated when the collapse proceeds from an initially static configuration. This investigation highlights the importance of the initial conditions during gravitational collapse.

\newpage

\begin{figure*}[thbp]
\begin{center}
\begin{tabular}{rl}
\includegraphics[width=5.6cm]{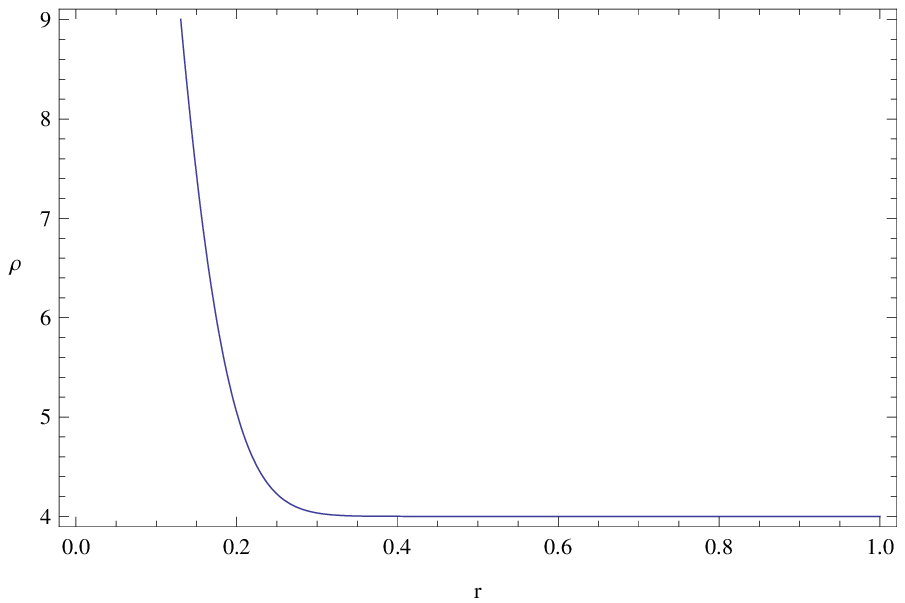}
\end{tabular}
\end{center}
\caption{Matter density $\rho$ is plotted against $r$ inside the fluid sphere by taking the values of the arbitrary constants $a=1$, $b=0.3$, $c=0.1$, $\mu=1$, $d=1$, $f=0.5$, $\alpha=1$ and $\beta=-10$ for the Wyman solution. }
\end{figure*}

\begin{figure*}[thbp]
\begin{center}
\begin{tabular}{rl}
\includegraphics[width=5.6cm]{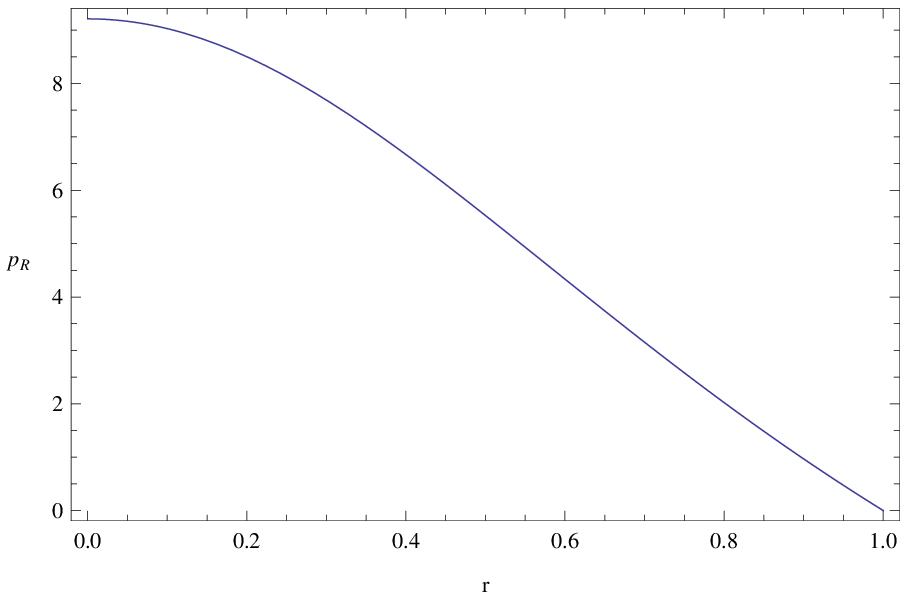}&
\includegraphics[width=5.6cm]{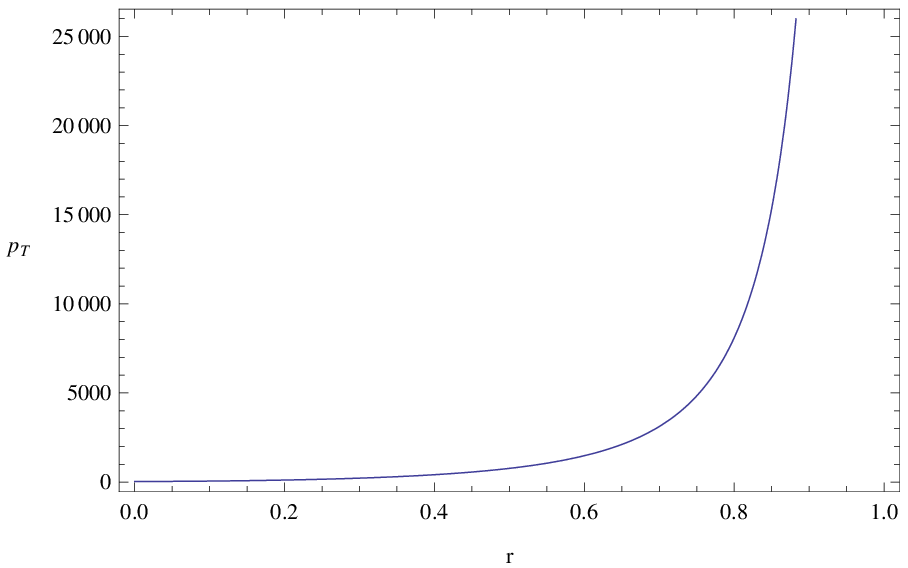}
\end{tabular}
\end{center}
\caption{The radial pressure $p_R$ (left panel) and transverse pressure $p_T$ (right panel) are plotted against r inside the stellar interior by employing the values of the arbitrary constants $a=1$, $b=0.3$, $c=0.1$, $\mu=1$, $d=1$, $f=0.5$, $\alpha=1$ and $\beta=1$ for $p_R$ and $a=1$, $b=0.3$, $c=1$, $\mu=1$, $d=1$, $f=0.5$, $\alpha=1$ and $\beta=-10$ for $p_T$ for the Wyman solution.}
\end{figure*}

\begin{figure*}[thbp]
\begin{center}
\begin{tabular}{rl}
\includegraphics[width=5.6cm]{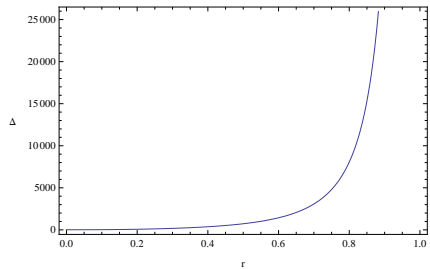}
\end{tabular}
\end{center}
\caption{The anisotropy parameter ($\Delta$) is plotted against the radial co-ordinate $r$ by employing the values of the arbitrary constants $a=1$, $b=1$, $c=1$, $\mu=1$, $d=1$, $f=0.5$, $\alpha=1$ and $\beta=-10$ for the Wyman solution. }
\end{figure*}

\begin{figure*}[thbp]
\begin{center}
\begin{tabular}{rl}
\includegraphics[width=5.6cm]{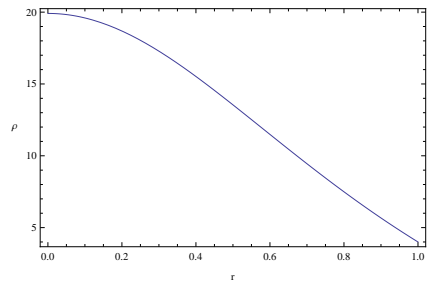}
\end{tabular}
\end{center}
\caption{Matter density $\rho$ is plotted against $r$ inside the fluid sphere by taking the values of the arbitrary constants $a=1$, $b=0.3$, $c=1$, $\mu=1$, $d=1$ and $f=0.5$ for the Florides solution. }
\end{figure*}

\begin{figure*}[thbp]
\begin{center}
\begin{tabular}{rl}
\includegraphics[width=5.6cm]{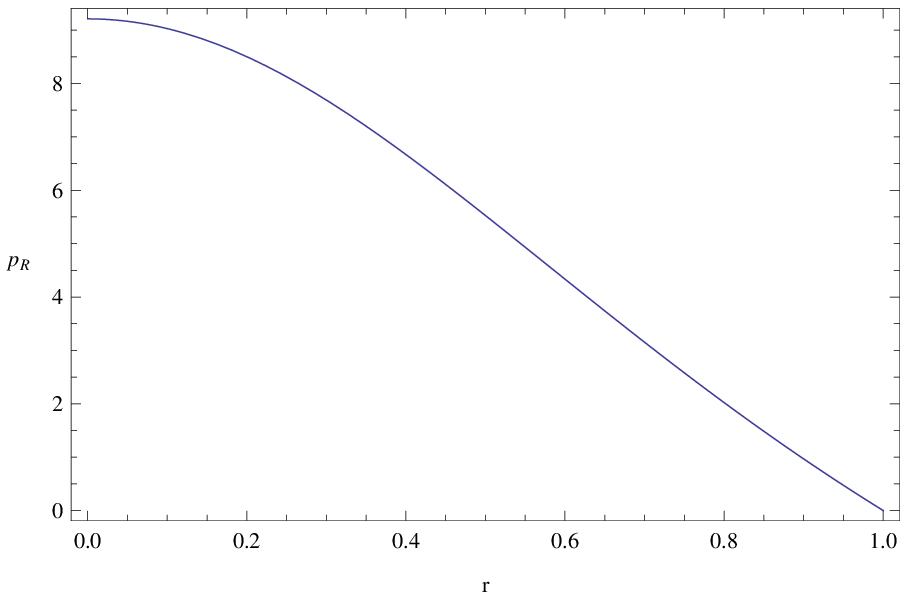}&
\includegraphics[width=5.6cm]{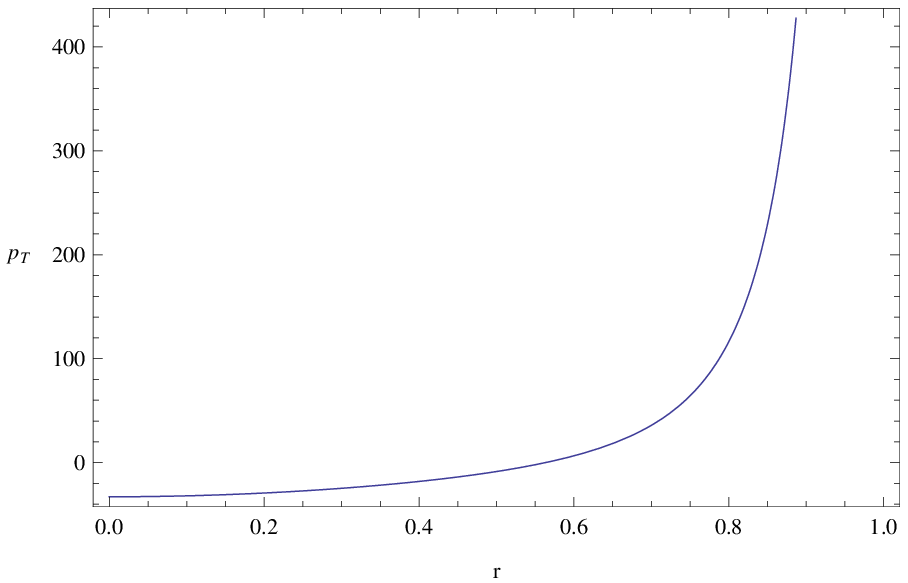}
\end{tabular}
\end{center}
\caption{The radial pressure $p_R$ (left panel) and transverse pressure $p_T$ (right panel) are plotted against $r$ inside the stellar interior by employing the values of the arbitrary constants $a=1$, $b=0.3$, $c=1$, $\mu=1$, $d=1$ and $f=0.5$ for $p_R$ and $a=1$, $b=0.3$, $c=1$, $\mu=1$, $d=1$ and $f=0.2$ for $p_T$ for the Florides solution.}
\end{figure*}

\begin{figure*}[thbp]
\begin{center}
\begin{tabular}{rl}
\includegraphics[width=5.6cm]{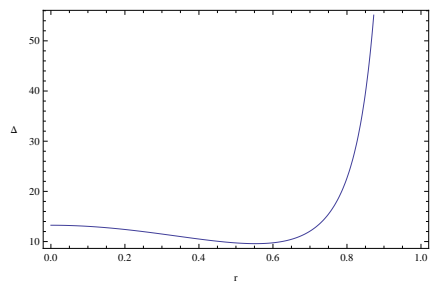}
\end{tabular}
\end{center}
\caption{The anisotropy parameter ($\Delta$)is plotted against the radial co-ordinate $r$ by employing the values of the arbitrary constants $a=1$, $b=1$, $c=1$, $\mu=1$, $d=1$ and $f=0.5$ for the Florides solution. }
\end{figure*}

\begin{figure*}[thbp]
\begin{center}
\begin{tabular}{rl}
\includegraphics[width=5.6cm]{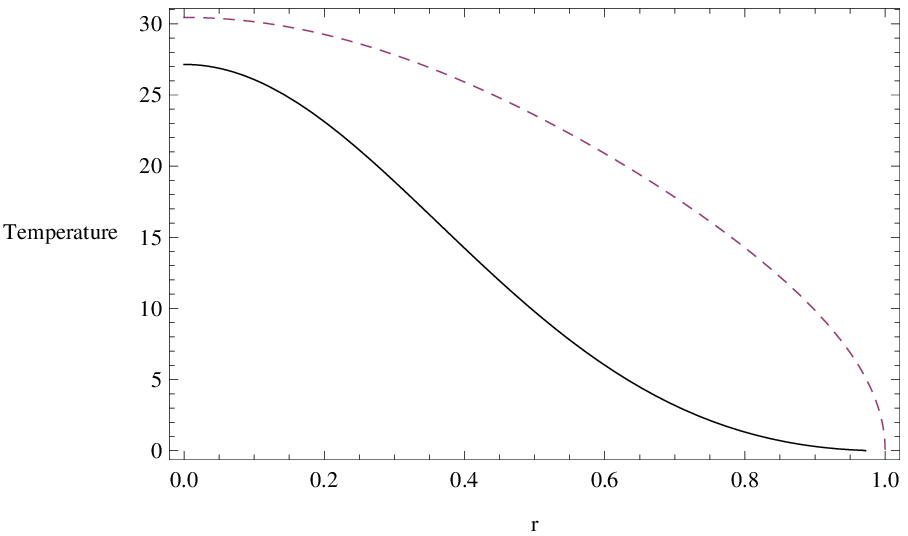}
\end{tabular}
\end{center}
\caption{The Eckart temperature is plotted against the radial coordinate $r$ for the Wyman (solid line) and Florides (dashed line)  models.}
\end{figure*}

\begin{figure*}[thbp]
\begin{center}
\begin{tabular}{rl}
\includegraphics[width=5.6cm]{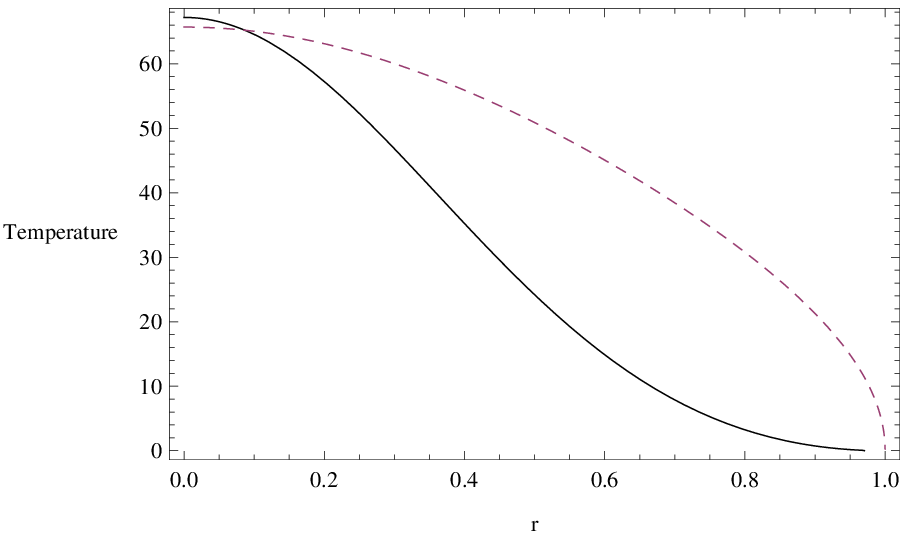}
\end{tabular}
\end{center}
\caption{The causal temperature is plotted against the radial coordinate $r$ for the Wyman (solid line) and Florides (dashed line) models.}
\end{figure*}

\end{document}